\documentclass[aps,prd,twocolumn,showpacs,amsmath,amssymb,floatfix]{revtex4-1}
\usepackage{graphicx,color,framed}
\usepackage{hyperref}
\usepackage{times}
\usepackage{enumerate}
\usepackage{lipsum}
\usepackage{slashed}
\usepackage{array}
\usepackage{textcomp}
\usepackage{amsthm}

\hypersetup{
    colorlinks=true, 
    linktoc=all,     
    linkcolor=blue,  
}

\newtheorem{thm}{Theorem}

\newtheorem{lemma}{Lemma}
\newtheorem{cor}{Corollary}

\def \beq {\begin{equation}}
\def \eeq {\end{equation}}
\def \beqa {\begin{eqnarray}}
\def \eeqa {\end{eqnarray}}
\def \bseq {\begin{subequations}}
\def \eseq {\end{subequations}}
\newcommand \dg {\dagger}
\newcommand \up {\uparrow}
\newcommand \down {\downarrow}

\newcommand \ran {\rangle}
\newcommand \lan {\langle}

\newcommand \ep {\epsilon}

\newcommand \mb {\mathbf}

\newcommand \nnb {\nonumber}
\newcommand \ov {\overline}
\newcommand \td {\tilde}

\begin{document}

\title{Topology of superconductors beyond mean-field theory}

\author{Matthew F. Lapa}
\email[email address: ]{mlapa@uchicago.edu}
\affiliation{Kadanoff Center for Theoretical Physics, University of Chicago, Chicago, Illinois 60637, USA}

\begin{abstract}

The study of topological superconductivity is largely based on the analysis of mean-field
Hamiltonians that violate particle number conservation and have only short-range interactions. Although
this approach has been very successful, it is not clear that it captures the topological properties of real 
superconductors, which are described by number-conserving Hamiltonians with long-range interactions.
To address this issue, we study topological superconductivity directly in the
number-conserving setting. We focus on a diagnostic for topological 
superconductivity that compares the fermion parity $\mathcal{P}$ of the ground state of a system in a ring 
geometry and in the presence of zero vs. $\Phi_{\text{sc}}=\frac{h}{2e} \equiv \pi$ flux of an external 
magnetic field. A version of this diagnostic exists in any dimension and provides a 
$\mathbb{Z}_2$ invariant $\nu=\mathcal{P}_0\mathcal{P}_{\pi}$ for topological superconductivity. 
In this paper we prove that the mean-field approximation correctly predicts the value of $\nu$ for a
large family of number-conserving models of spinless superconductors. 
Our result applies directly to the cases of 
greatest physical interest, including $p$-wave and $p_x+ip_y$ superconductors
in one and two dimensions, and gives strong evidence for the validity of the mean-field approximation 
in the study of (at least some aspects of) topological superconductivity.

\end{abstract}


\maketitle

\section{Introduction} 

Topological superconductors 
(TSCs)~\cite{kitaev,read-green,tanaka2011symmetry,alicea2012new,sato-review} 
are of great interest both from a theoretical point of
view and for their possible applications to quantum computation. However, most theoretical 
studies of TSCs rely on simple mean-field Hamiltonians that violate particle number conservation, and over 
the past few years several authors have expressed concerns about this 
approach~\cite{Leggett1,Leggett2,ortiz2,lin-leggett1,lin-leggett2}. 
In addition, on the experimental side, the interpretation of 
transport measurements designed to search for Majorana fermions is not yet 
clear~\cite{zb1,zb2,ji-wen2018,sau2018,absence2020}. 
To gain a better understanding of these issues, in Ref.~\onlinecite{LL} we initiated 
a rigorous study of TSCs in the more realistic number-conserving setting. 

A key theoretical concern highlighted in Ref.~\onlinecite{LL} is the following. 
In the number-conserving setting one must include \emph{long-range}
interactions to accurately describe real charged superconductors 
(e.g., to avoid the Goldstone theorem of Ref.~\onlinecite{Hastings1}). However, it
is not clear that the topological properties
of number-conserving superconductor models with long-range interactions
can be correctly captured by mean-field models that violate particle number conservation
and have only short-range interactions.

In this paper we address this concern by studying the topological phases of 
superconductors beyond mean-field theory. Specifically, we study a diagnostic 
for topological superconductivity that compares
the fermion parity $\mathcal{P}$ of the ground state~\footnote{For a model
with a conserved total particle number, by ``ground state'' we mean the lowest energy state over 
all possible particle number sectors.} of a system in a ring geometry and
in the presence of zero vs. $\pi$ flux of an external magnetic field~\cite{kitaev,ortiz1,ortiz2}. 
Here, $\pi$ flux is equivalent to one superconducting flux quantum $\Phi_{\text{sc}}=\frac{h}{2e}$ in units 
with $\hbar=e=1$. A version of this diagnostic exists in any spatial dimension and
provides a $\mathbb{Z}_2$ invariant 
\beq
	\nu= \mathcal{P}_0\mathcal{P}_{\pi}
\eeq
for topological superconductivity. The nontrivial phase corresponds to $\nu=-1$, where a ``fermion parity 
switch'' occurs between zero and $\pi$ flux. More physically, for mean-field models
the value $\nu=-1$ indicates that gapless \emph{Majorana} excitations 
appear at the boundaries of a finite system.
The invariant $\nu$ is only well-defined if the Hamiltonian at zero and $\pi$ 
flux possesses a finite \emph{parity gap}, i.e., a finite energy gap between the ground state and the
lowest energy state with opposite parity. Note also that $\nu$ only makes sense in the thermodynamic
limit if the parity gaps remain finite in that limit.

We study the invariant $\nu$ in a general family of number-conserving \emph{pairing} models of 
spinless fermions in $D$ spatial dimensions. We also restrict our attention to 
translation invariant models, which allows us to apply the flux via a change in the boundary 
conditions. These pairing models are similar to the reduced BCS model~\cite{BCS} and to 
Richardson-type models~\cite{richardson1,richardson2}, and they form a convenient starting point for the 
study of superconductivity in the number-conserving setting.  
In special cases exactly solvable versions of these models have already been used to obtain 
detailed results on $p$-wave superconductors in 
$D=1$ and $p_x+ip_y$ superconductors (and more exotic cases) in 
$D=2$~\cite{ortiz1,ortiz2,sierra1,sierra2,RDO,lwave}. In addition, Ref.~\onlinecite{ortiz1} 
proved that $\nu=-1$ in an exactly solvable model in $D=1$~\footnote{In this paper we define
$\nu$ using the parity of the lowest energy state of a given Hamiltonian over all possible
particle number sectors. This should be contrasted
with the approach in Ref.~\onlinecite{ortiz1}, where the authors considered a fixed 
$N$ and checked whether the sign of $E_0^{(N)}-\frac{1}{2}(E_0^{(N-1)}+E_0^{(N+1)})$ was different for the
two choices of boundary condition (here, $E_0^{(N)}$ is the ground state energy in the $N$-particle sector).}.

In this paper we prove that for gapped pairing models $\nu$ satisfies the relation
\beq
	\nu= \prod_{\mb{k}\in\mathcal{K}_0}s_{\mb{k}}\ , \label{eq:main-result}
\eeq
where $s_{\mb{k}}= \text{sgn}(\ep_{\mb{k}})$, $\ep_{\mb{k}}$ is the single particle energy
dispersion in the pairing model, and $\mathcal{K}_0$ is the set of time-reversal invariant wave 
vectors in the first Brillouin zone (these satisfy $\mb{k}=-\mb{k}+\mb{G}$ for some
reciprocal lattice vector $\mb{G}$). This is exactly the result one would obtain
for $\nu$ by studying the pairing model using a BCS-type mean-field 
approximation, which reduces the pairing model to a quadratic mean-field model. In addition,
the product $\prod_{\mb{k}\in\mathcal{K}_0}s_{\mb{k}}$ is known to be a 
$\mathbb{Z}_2$ topological invariant for mean-field models of spinless superconductors in 
$D=1$ and $2$~\cite{kitaev,sato1}. 
Therefore our result proves that the mean-field approximation correctly predicts the value of 
$\nu$ for these gapped pairing models and, by the definition of $\nu$, 
for \emph{any} model that is adiabatically connected to a gapped pairing model. 
For the precise statements of these results, see Theorem 1 and Corollary 1 below.

Previous studies of TSCs with number conservation have mostly focused on
one-dimensional systems~\cite{cheng-tu,
fidkowski2011,Sau2011,tsvelik2011zero,cheng-lutchyn,KSH,knapp2019number,
ortiz1,ortiz2,iemini2015localized,lang2015topological,
PhysRevB.96.115110,Kraus2013,burnell2018,RA,yin2019majorana,LL}. 
Important exceptions include Refs.~\onlinecite{sierra1,sierra2,RDO,lwave}, 
which considered exactly solvable pairing models in $D=2$, and 
Refs.~\onlinecite{lin-leggett1,lin-leggett2}, 
which considered the braiding statistics of vortices in $D=2$. Our 
rigorous results for all $D\geq 1$ should nicely complement these previous studies and serve as a useful 
guide for future work on this topic. 

This paper is organized as follows. In Sec.~\ref{sec:models} we introduce the family of number-conserving 
pairing models that we study in this paper. In Sec.~\ref{sec:results} we present our main results, 
and we also state and prove a lemma that is used in the proof of our main results.
In Sec.~\ref{sec:thm1-proof} we present the proof of our main result 
(namely, Theorem 1 from Sec.~\ref{sec:results}). Section~\ref{sec:con} presents our conclusions.
Finally, Appendix~\ref{app:RP} contains an introduction to the \emph{reflection positivity} property that
is used our proof of Lemma 1 in Sec.~\ref{sec:results}.

\section{Number-conserving pairing models of spinless fermions} 
\label{sec:models}

In this section we introduce the
pairing models that we study in this paper. The degrees of freedom in these models are 
spinless fermions $c_{\mb{x}}$ on the sites
$\mb{x}$ of a Bravais lattice $\Lambda$ in $D$ spatial dimensions ($D\geq 1$), 
and these operators obey standard 
anti-commutation relations $\{c_{\mb{x}},c_{\mb{y}}\}=0$ and 
$\{c_{\mb{x}},c^{\dg}_{\mb{y}}\}=\delta_{\mb{x}\mb{y}}$. To avoid unnecessary complications, we also assume 
that the number of unit cells in the lattice in each coordinate direction is an even number.

We consider translation invariant models
with two different boundary conditions. In the first case, we consider periodic boundary
conditions in all coordinate directions, corresponding to the absence of any magnetic flux. 
In the second case, we consider anti-periodic boundary conditions in a \emph{single} coordinate direction, 
and periodic boundary conditions in the remaining $D-1$ directions. This corresponds to the 
presence of $\Phi_{\text{sc}}=\frac{h}{2e} \equiv \pi$ flux through a single hole in the 
$D$-dimensional torus on which our system lives. In $D=1$ the second case just
corresponds to standard anti-periodic boundary conditions. 
In each case the boundary conditions determine a
set $\mathcal{K}$ of allowed wave vectors in the first Brillouin zone (BZ). For each
allowed wave vector $\mb{k}$ we can define a fermion operator in reciprocal space via 
Fourier transform, $c_{\mb{k}} = |\Lambda|^{-1/2}\sum_{\mb{x}} c_{\mb{x}} e^{-i\mb{k}\cdot\mb{x}}$, where
$|\Lambda|$ is the number of unit cells in the lattice.

We can always decompose $\mathcal{K}$ as
$\mathcal{K}=\mathcal{K}_0\cup \mathcal{K}_{+}\cup\mathcal{K}_{-}$, where the three factors here
are as follows. The first set $\mathcal{K}_0$ consists of all time-reversal invariant
wave vectors in the first BZ. These wave vectors satisfy $\mb{k}=-\mb{k}+\mb{G}$ for some reciprocal lattice 
vector $\mb{G}$. The remaining 
factor $\mathcal{K}_{+}\cup\mathcal{K}_{-}$ denotes \emph{any} decomposition of the remaining wave
vectors into two sets in such a way that, if $\mb{k}\in\mathcal{K}_{+}$, then $-\mb{k}\in\mathcal{K}_{-}$. 
The significance of the set $\mathcal{K}_0$ is that fermions at the wave vectors in this set do not 
participate in the pairing interaction in the Hamiltonians that we consider. A crucial point
for the remainder of the paper is that in the case of anti-periodic boundary conditions 
we have $\mathcal{K}_0=\emptyset$, i.e., there are no
time-reversal invariant wave vectors in the first BZ in the anti-periodic case.

It is helpful to illustrate our notation with an example. 
Consider a one-dimensional system in a ring geometry with
$L$ sites and $L$ even (and with a lattice spacing equal to 1). 
Then in the periodic case we have
$\mathcal{K}_0=\{0,\pi\}$ and 
$\mathcal{K}_{+}=\left\{\tfrac{2\pi}{L},\tfrac{4\pi}{L},\dots,\pi-\tfrac{2\pi}{L}\right\}$, while in the
anti-periodic case we have $\mathcal{K}_0=\emptyset$ and 
$\mathcal{K}_{+}=\left\{\tfrac{\pi}{L},\tfrac{3\pi}{L},\dots,\pi-\tfrac{\pi}{L}\right\}$.

The Hamiltonian for the pairing models that we consider, for either choice of boundary
condition, takes the general form
\beq
	H= \sum_{\mb{k}\in\mathcal{K}} \ep_{\mb{k}} n_{\mb{k}} - \sum_{\mb{k},\mb{k}'\in \mathcal{K}_{+}} g_{\mb{k}\mb{k}'} c^{\dg}_{\mb{k}} c^{\dg}_{-\mb{k}}c_{-\mb{k}'}c_{\mb{k}'}\ , \label{eq:Ham}
\eeq
where $n_{\mb{k}}=c^{\dg}_{\mb{k}} c_{\mb{k}}$, $\ep_{\mb{k}}$ is a single particle energy dispersion, and 
$g_{\mb{k}\mb{k}'}$ parametrizes the interaction between the pairs at $(\mb{k},-\mb{k})$ and 
$(\mb{k}',-\mb{k}')$. Note that $H$ commutes with the total particle number operator 
$\mathcal{N}=\sum_{\mb{k}}c^{\dg}_{\mb{k}} c_{\mb{k}}=\sum_{\mb{x}}c^{\dg}_{\mb{x}} c_{\mb{x}}$. 
We also absorb any chemical potential term $-\mu\mathcal{N}$ into the definition of $\ep_{\mb{k}}$.

We make the following assumptions about $\ep_{\mb{k}}$ and $g_{\mb{k}\mb{k}'}$. 
First, we assume that $\ep_{\mb{k}}$ is an even function of $\mb{k}$ for $\mb{k}\notin\mathcal{K}_0$,
\beq
	\ep_{\mb{k}}=\ep_{-\mb{k}}\ \ \forall\ \ \mb{k}\in\mathcal{K}_{+}\ .
\eeq
Next, we assume that $g_{\mb{k}\mb{k}'}$ takes the factorized form
\beq
	g_{\mb{k}\mb{k}'}= \ov{\eta}_{\mb{k}}\eta_{\mb{k}'}\ , \label{eq:factor}
\eeq
where $\eta_{\mb{k}}$ is a complex function of $\mb{k}$ (the overline denotes complex conjugation),
and we assume that $\eta_{\mb{k}}\neq 0$ for all $\mb{k}\in\mathcal{K}_{+}$.
For specific examples of models of this form, which include the cases of $p$-wave superconductors in 
$D=1$ and $p_x+ip_y$ superconductors in $D=2$, we refer the reader to 
Refs.~\onlinecite{ortiz1,ortiz2,sierra1,sierra2,RDO,lwave}. We note here that, unlike those references,
we do not assume any fine-tuning of $\ep_{\mb{k}}$ or $\eta_{\mb{k}}$ that might lead to exact solvability.

One benefit of the factorization assumption \eqref{eq:factor} is that it implies that these pairing models
also take a sensible form in real space. In this case each individual sum
$\sum_{\mb{k}\in \mathcal{K}_{+}} \eta_{\mb{k}} c_{-\mb{k}} c_{\mb{k}}$ in the pairing term can be
Fourier transformed back to real space, and one finds that in real space the pairing term becomes a 
long-range pair hopping term. 

Finally, let $\Delta_{0}$ and $\Delta_{\pi}$ be the parity gaps of the 
Hamiltonian with the two choices of boundary condition/magnetic flux. The invariant $\nu$ is only 
well-defined if both of these gaps are non-zero. In the periodic case we trivially find that $\Delta_0=0$ 
if $\ep_{\mb{k}}=0$ for any $\mb{k}\in\mathcal{K}_0$, and so in what follows we always assume that 
$\ep_{\mb{k}}\neq 0$ for all $\mb{k}\in\mathcal{K}_0$.

\section{Main results}
\label{sec:results}

In this section we present our main results (Theorem 1 and Corollary 1), and we also state and prove a lemma 
(Lemma 1) that is used in the proof of our main results. Our first result is a formula for $\nu$ in gapped 
pairing models.

\begin{thm}
Let $H$ be a pairing Hamiltonian of the form \eqref{eq:Ham} with non-zero parity gaps
$\Delta_{0}$ and $\Delta_{\pi}$. Then for this Hamiltonian $\nu$ satisfies the relation
\beq
	\nu= \prod_{\mb{k}\in\mathcal{K}_0}s_{\mb{k}}\ , \label{eq:I-formula}
\eeq
where $s_{\mb{k}}= \text{sgn}(\ep_{\mb{k}})$ and $\mathcal{K}_0$
is the set of time-reversal invariant wave vectors in the first Brillouin zone for the case of
periodic boundary conditions.
\end{thm}

By combining Theorem 1 with the definition of $\mathcal{P}_0$ and $\mathcal{P}_{\pi}$, we immediately obtain 
the following corollary. 

\begin{cor}
Let $H_0$ be a pairing Hamiltonian for which Theorem 1 applies, and let $H_1$ be 
any other translation invariant Hamiltonian such that $H_s = (1-s)H_0 + sH_1$ has non-zero parity 
gaps $\Delta_0(s)$ and $\Delta_{\pi}(s)$ for all $s\in[0,1]$. Then
\beq
	\nu_1= \nu_0= \prod_{\mb{k}\in\mathcal{K}_0}s_{\mb{k}}\ ,
\eeq
where $\nu_0$ and $\nu_1$ are the invariants for $H_0$ and $H_1$, 
respectively.
\end{cor}

Theorem 1 shows that for gapped pairing models the invariant $\nu$ is equal to the value
that one would predict using a BCS-type mean-field approximation, namely the value
$\prod_{\mb{k}\in\mathcal{K}_0}s_{\mb{k}}$. The product $\prod_{\mb{k}\in\mathcal{K}_0}s_{\mb{k}}$ 
is known to be a $\mathbb{Z}_2$ topological invariant for quadratic mean-field models of 
spinless superconductors, and in the mean-field context it was originally derived in 
Ref.~\onlinecite{kitaev} for the case of $D=1$ and Ref.~\onlinecite{sato1} for $D=2$ 
(see also Refs.~\onlinecite{sato2,fu-berg} for $D>1$).
Therefore, Theorem 1 and Corollary 1 prove that the mean-field approximation correctly 
predicts the value of $\nu$ for \emph{any} translation invariant model of spinless fermions 
that is adiabatically connected to a gapped pairing model. 
Note that for Corollary 1 we do not need
to assume that $H_1$ is number-conserving, but here we do assume that $H_1$ is translation invariant so 
that we can apply the $\pi$ flux via a change in the boundary conditions (i.e., by the choice of 
the set $\mathcal{K}$ of allowed wave vectors).

One possible application of Corollary 1 is to predict a topological superconducting phase
in realistic Hamiltonians. For example, $H_1$ might be a Hamiltonian of the form
\beq
	H_1= \sum_{\mb{k}\in\mathcal{K}} \ep_{\mb{k}} n_{\mb{k}} + \sum_{\mb{x},\mb{y}\in\Lambda}v_{\mb{x}\mb{y}}n_{\mb{x}}n_{\mb{y}}\ ,
\eeq	
where $v_{\mb{x}\mb{y}}$ is a translation invariant density-density interaction in 
real space ($n_{\mb{x}}=c^{\dg}_{\mb{x}}c_{\mb{x}}$). If 
$v_{\mb{x}\mb{y}}$ favors a superconducting ground state with a finite parity gap then, 
following the logic of the original BCS paper~\cite{BCS}, it is possible that $H_1$ is 
adiabatically connected to a gapped pairing Hamiltonian of the form \eqref{eq:Ham}. In that case
we could then use Corollary 1 to predict whether $H_1$ supports a topological superconducting phase.

To prove Theorem 1 we first rewrite $H$ in the form
$H= \sum_{\mb{k}\in\mathcal{K}_0} \ep_{\mb{k}} n_{\mb{k}} + \tilde{H}$,
where
\beq
	\tilde{H}= \sum_{\mb{k}\in\mathcal{K}_{+}\cup\mathcal{K}_{-}} \ep_{\mb{k}} n_{\mb{k}} - \sum_{\mb{k},\mb{k}'\in \mathcal{K}_{+}} g_{\mb{k}\mb{k}'} c^{\dg}_{\mb{k}} c^{\dg}_{-\mb{k}}c_{-\mb{k}'}c_{\mb{k}'}\ . \label{eq:td-Ham}
\eeq
Note that $\tilde{H}$ only contains the fermions that participate
in the pairing interaction.
The proof of Theorem 1 relies on a lemma (Lemma 1) that
concerns the ground state of $\tilde{H}$ acting within the space 
$\tilde{\mathcal{F}}$ consisting of those states annihilated by
all the $c_{\mb{k}}$ with $\mb{k}\in\mathcal{K}_0$, 
\beq
	\tilde{\mathcal{F}}= \left\{|\psi\ran\  :\  c_{\mb{k}}|\psi\ran = 0 \ \forall\ \mb{k}\in\mathcal{K}_0\right\}\ .  \label{eq:Fock}
\eeq

Let $\tilde{E}^{(M)}_0$ be the ground state 
energy of $\tilde{H}$ in the $M$-particle sector of $\tilde{\mathcal{F}}$, and
let $|\tilde{\psi}^{(M)}_0\ran$ be the corresponding ground state (or a particular ground state if 
$\tilde{H}$ has a ground state degeneracy in that sector). Note that $M$ must satisfy 
$0\leq M\leq |\Lambda|-|\mathcal{K}_0|$ (where $|\mathcal{K}_0|$ is the number of wave vectors in
$\mathcal{K}_0$) since we are working within the space $\tilde{\mathcal{F}}$.
Finally, let $M^*$, with $0\leq M^*\leq |\Lambda|-|\mathcal{K}_0|$, be the (not necessarily unique)
integer satisfying
\beq
	\tilde{E}^{(M)}_0 \geq \tilde{E}^{(M^*)}_0 \ \ \forall\ \ M\ \ \text{with}\ 0\leq M\leq |\Lambda|-|\mathcal{K}_0|\ .
\eeq
Thus, $M^*$ (if it is unique) is the particle number in the ground state of $\tilde{H}$ acting in the 
space $\tilde{\mathcal{F}}$.
The integer $M^*$ plays an important role in our proof below, and for this integer we have
the following result.

\begin{lemma}
The integer $M^*$ can always be chosen to be \emph{even}.
\end{lemma}

\subsection{Proof of Lemma 1} 

We consider $\tilde{H}$ acting in the
space $\tilde{\mathcal{F}}$. The proof is based on the fact that 
$\tilde{H}$ possesses the property of \emph{reflection positivity}~\cite{KLS,lieb-hubbard,LN}.
For the reader's benefit, we provide an introduction to this property in Appendix~\ref{app:RP}. 
In our case the ``reflection'' one needs to consider is actually
\emph{inversion} in momentum space, $\mb{k}\to-\mb{k}$. 
However, we show below that $\tilde{H}$ can be mapped exactly to a Hamiltonian possessing 
Lieb's \emph{spin reflection positivity}~\cite{lieb-hubbard}. 
We can then immediately apply the results of Tian and Tang~\cite{TT}
on \emph{spinful} pairing models to prove Lemma 1. 

The Hamiltonian $\tilde{H}$ is written in terms of spinless fermions labeled by wave vectors 
$\mb{k}$ in both sets $\mathcal{K}_{+}$ and $\mathcal{K}_{-}$. We now perform a change of variables to 
``spinful'' fermions  $c_{\mb{k},\sigma}$, $\sigma\in\{\up,\down\}$,
labeled by wave vectors in $\mathcal{K}_{+}$ only. To 
define these new variables we first decompose $\eta_{\mb{k}}$ into magnitude and phase parts as
$\eta_{\mb{k}}=|\eta_{\mb{k}}|e^{i\theta_{\mb{k}}}$. 
Then, for any $\mb{k}\in\mathcal{K}_{+}$ we define 
\begin{subequations}
\beqa
	c_{\mb{k},\up} &=& c_{\mb{k}} \\
	c_{\mb{k},\down} &=& e^{i\theta_{\mb{k}}} c_{-\mb{k}}\ ,
\eeqa
\end{subequations}
and one can check that these operators obey standard anti-commutation relations for spinful
fermions. In terms of these new operators $\tilde{H}$ can be written in the form
\beq
	\tilde{H}= \sum_{\mb{k}\in\mathcal{K}_{+}}\sum_{\sigma=\up,\down}\ep_{\mb{k}} n_{\mb{k},\sigma} - \sum_{\mb{k},\mb{k}'\in \mathcal{K}_{+}} |\eta_{\mb{k}}||\eta_{\mb{k}'}| c^{\dg}_{\mb{k},\up} c_{\mb{k}',\up} c^{\dg}_{\mb{k},\down}c_{\mb{k}',\down}\ ,
\eeq
where $n_{\mb{k},\sigma}=c^{\dg}_{\mb{k},\sigma}c_{\mb{k},\sigma}$, and where we used
$\ep_{\mb{k}}=\ep_{-\mb{k}}$ for $\mb{k}\in\mathcal{K}_{+}$
and also rearranged the order of the operators in the pairing term.
With this change of variables
$\tilde{H}$ now has the form of the 
pairing Hamiltonians for spinful fermions studied in Ref.~\onlinecite{TT} (see their Eq.~5), and so
it possesses Lieb's spin reflection positivity. We can then
apply the results of Ref.~\onlinecite{TT} (specifically, their Eq.~26) to conclude that 
\beqa
	\tilde{E}^{(2M+1)}_0 &\geq& \frac{1}{2}\left( \tilde{E}^{(2M)}_0 + \tilde{E}^{(2M+2)}_0 \right) \nnb \\
	&\geq& \min\left\{\tilde{E}^{(2M)}_0,\ \tilde{E}^{(2M+2)}_0 \right\}\ , \label{eq:RP-result}
\eeqa
which proves Lemma 1.

\section{Proof of Theorem 1}
\label{sec:thm1-proof}

In this section we present the proof of Theorem 1. 
We first prove the theorem in the case of $D=1$.
The proof for $D> 1$ is conceptually identical to the $D=1$ case, but requires more cumbersome notation, 
and so we present the proof for $D>1$ after the proof for $D=1$.

\subsection{One-dimensional case} 

For the case of $D=1$
we take $\Lambda$ to be a linear chain with $L$ sites and $L$ even (and with a lattice spacing equal to 1). 
With this choice we have
$\mathcal{K}_0= \{0,\pi\}$ for periodic boundary conditions (zero flux), while $\mathcal{K}_0=\emptyset$ 
for anti-periodic boundary conditions ($\pi$ flux).

In the anti-periodic case, Lemma 1 and our assumption of a non-zero parity gap $\Delta_{\pi}$ immediately
imply that $\mathcal{P}_{\pi}=1$. Therefore it remains to compute $\mathcal{P}_0$. 
Our strategy for this is as follows. Let $E^{(N)}_0$ be the ground state energy
of $H$ in the $N$-particle sector. The first part of the proof consists of establishing (for any
$N$) a lower bound on $E^{(N)}_0$ of the form $E^{(N)}_0 \geq E^{(M^{**})}_0$, where 
$M^{**}$ is a particular integer that we define below. Using our
assumption of a non-zero parity gap $\Delta_0$, this implies that 
$\mathcal{P}_0=(-1)^{M^{**}}$. Finally, we use Lemma 1 to prove that the integer $M^{**}$ satisfies the
relation $(-1)^{M^{**}}= s_0s_{\pi}$, which completes the proof of Theorem 1.

To start, let $|\psi^{(N)}_0\ran$ be the ground state of $H$ in the
$N$-particle sector (or one of the ground states if there is a degeneracy in that sector). 
We can always write $|\psi^{(N)}_0\ran$ in the form
\beq
	|\psi^{(N)}_0\ran= \sum_{b_1,b_2=0,1}a_{b_1 b_2}(c^{\dg}_0)^{b_1}( c^{\dg}_{\pi})^{b_2}
	|\psi^{(N-b_1-b_2)}_{b_1b_2}\ran\ ,
\eeq
where the four states $|\psi^{(N)}_{00}\ran, |\psi^{(N-1)}_{10}\ran, |\psi^{(N-1)}_{01}\ran$, and 
$|\psi^{(N-2)}_{11}\ran$ are all annihilated by $c_0$ and $c_{\pi}$ (i.e., they
are states in $\tilde{\mathcal{F}}$), 
and where the coefficients
$a_{b_1b_2}$ satisfy $\sum_{b_1,b_2=0,1}|a_{b_1 b_2}|^2=1$. Note, however, that if $N=L-1$ then we must have
$a_{00}=0$, and if $N=L$ then we must have $a_{00}=a_{10}=a_{01}=0$. On the other hand, if $N\leq L-2$, then
all of $a_{b_1b_2}$ can be non-zero in general. In what follows we consider the
case where $N\leq L-2$ so that $a_{b_1b_2}\neq 0$ in general, but all of our
results also hold for $N=L-1$ and $N=L$ and can be derived in the same way (but setting
$a_{00}=0$ or $a_{00}=a_{10}=a_{01}=0$ from the start for the two cases of $N=L-1$ and $N=L$, 
respectively).

Using the fact that $[H,n_0]=[H,n_{\pi}]=0$, 
we find that $E^{(N)}_0= \lan \psi^{(N)}_0|H|\psi^{(N)}_0\ran$ takes the form
\begin{align}
	E^{(N)}_0 &= |a_{00}|^2 \lan \psi^{(N)}_{00}|\td{H}|\psi^{(N)}_{00}\ran \nnb \\
	&+ |a_{10}|^2\left(\ep_0 + 
	\lan \psi^{(N-1)}_{10}|\td{H}|\psi^{(N-1)}_{10}\ran\right) \nnb \\
	&+ |a_{01}|^2\left(\ep_{\pi} + 
	\lan \psi^{(N-1)}_{01}|\td{H}|\psi^{(N-1)}_{01}\ran\right) \nnb \\
	&+ |a_{11}|^2\left(\ep_0 + \ep_{\pi}+
	\lan \psi^{(N-2)}_{11}|\td{H}|\psi^{(N-2)}_{11}\ran\right)\ .
\end{align}
Next, using the variational theorem for the ground state of $\tilde{H}$ restricted to
$\tilde{\mathcal{F}}$ (e.g.,
$\lan \psi^{(N)}_{00}|\td{H}|\psi^{(N)}_{00}\ran\geq \td{E}^{(N)}_0$), the fact that
$\tilde{E}^{(M)}_0 \geq \tilde{E}^{(M^*)}_0$ for any $M\leq L-2$, and
$\sum_{b_1,b_2=0,1}|a_{b_1 b_2}|^2=1$, we find that
\begin{align}
	E^{(N)}_0 \geq \tilde{E}^{(M^*)}_0+ (|a_{10}|^2+|a_{11}|^2)\ep_0 +(|a_{01}|^2+|a_{11}|^2)\ep_{\pi}\ . \label{eq:basic-ineq}
\end{align}
To proceed further, we define $h_0=(1-s_0)/2$ and 
$h_{\pi}=(1-s_{\pi})/2$, and note that $h_0\in\{0,1\}$ and likewise for $h_{\pi}$.  
The inequality \eqref{eq:basic-ineq}, combined with the constraint $\sum_{b_1,b_2=0,1}|a_{b_1 b_2}|^2=1$,
then implies that
\beq
	E^{(N)}_0 \geq h_0 \ep_0 + h_{\pi}\ep_{\pi}+\tilde{E}^{(M^*)}_0 \ . \label{eq:ineq1}
\eeq 
This inequality holds for any value of $N$, although in our derivation here
we considered the case of $N\leq L-2$.

The next step of the proof is to study the trial state
\beq
	|\psi_t\ran= (c^{\dg}_0)^{h_0}(c^{\dg}_{\pi})^{h_{\pi}}|\td{\psi}^{(M^*)}_0\ran\ ,
\eeq
where $|\td{\psi}^{(M^*)}_0\ran$ is the ground state of $\tilde{H}$ 
in the $M^*$-particle sector of $\tilde{\mathcal{F}}$ 
(or one of the ground states if $\td{H}$ has a degeneracy in that sector).
This trial state has a particle number equal to
\begin{align}
	M^{**}= M^* + h_0 + h_{\pi}\ .\label{eq:Mss}
\end{align}
In addition, the energy of this trial state is 
\beq
	E_t= \lan\psi_t|H|\psi_t\ran= h_0\ep_0 + h_{\pi}\ep_{\pi} + \tilde{E}^{(M^*)}_0\ .
\eeq
From this we can see that the inequality \eqref{eq:ineq1} can
be rewritten in the form
\beq
	E^{(N)}_0 \geq E_t \ \ \forall\ N\ . \label{eq:lower-bound}
\eeq
On the other hand, using the variational theorem for $H$ in the $M^{**}$-particle sector, we have the 
upper bound $E^{(M^{**})}_0 \leq E_t$.
Combining this upper bound with the lower bound \eqref{eq:lower-bound} yields an
\emph{equality} for the ground state energy in the $M^{**}$-particle sector,
$E^{(M^{**})}_0 = E_t$.
Then our previous inequality \eqref{eq:lower-bound} can be rewritten as
\beq
	E^{(N)}_0 \geq E^{(M^{**})}_0\ \ \forall\ N\ .
\eeq
This inequality shows that the ground state energy in \emph{any} sector of fixed particle number is 
greater than or equal to the ground state energy in the $M^{**}$-particle sector. By 
combining this inequality with our assumption of a non-zero parity gap $\Delta_0$, we find that
$\mathcal{P}_0=(-1)^{M^{**}}$.

Finally, we come to the crucial point. Using Lemma 1, which implies that we can take $M^*$ to be even,
we find that 
\beq
	(-1)^{M^{**}}= (-1)^{h_0+h_{\pi}}= s_0s_{\pi}\ .
\eeq
This completes the proof of Theorem 1 for the case of $D=1$.

\subsection{Higher dimensions}

In this section we prove Theorem 1 in any spatial dimension $D\geq 1$. The logic of the proof is exactly
the same as in the $D=1$ case from the previous subsection. To start, since $\mathcal{K}_0=\emptyset$ in the
anti-periodic case ($\pi$ flux), our assumption of a non-zero parity gap $\Delta_{\pi}$ again implies that
$\mathcal{P}_{\pi}=1$. So all that remains is to again calculate $\mathcal{P}_0$. 

We now introduce some notation that will streamline the calculation of $\mathcal{P}_0$ in this
higher-dimensional case. Let $H$ be the pairing Hamiltonian with periodic boundary conditions,
let $E^{(N)}_0$ be the ground state energy of $H$ in the $N$-particle sector, and let
$|\psi^{(N)}_0\ran$ be the ground state of $H$ in the $N$-particle sector (or one of the ground states if
$H$ has a degeneracy in that sector). In addition, let $\mb{k}_j$, 
for $j\in\{1,\dots,|\mathcal{K}_0|\}$, be the wave vectors in the
set $\mathcal{K}_0$, and let $s_j= \text{sgn}(\ep_{\mb{k}_j})$ and
$h_j= (1-s_j)/2$ (and recall that we assume that $\ep_{\mb{k}}\neq 0$ for 
all $\mb{k}\in\mathcal{K}_0$ to avoid a trivial vanishing of the parity gap $\Delta_0$).

To start, we note that we can
again write $|\psi^{(N)}_0\ran$ as a linear combination of states
with different occupations of the fermions labeled by wave vectors in $\mathcal{K}_0$. In
particular, we can write
\begin{widetext}
\begin{align}
	|\psi^{(N)}_0\ran = \sum_{b_1,\dots,b_{|\mathcal{K}_0|}=0,1}a_{b_1\cdots b_{|\mathcal{K}_0|}} 
	(c^{\dg}_{\mb{k}_1})^{b_1}\cdots (c^{\dg}_{\mb{k}_{|\mathcal{K}_0|}})^{b_{|\mathcal{K}_0|}}\big|\psi^{(N-\sum_{j=1}^{|\mathcal{K}_0|}b_j)}_{b_1\cdots b_{|\mathcal{K}_0|}}\big\ran\ ,
\end{align}
\end{widetext}
where the $2^{|\mathcal{K}_0|}$ states 
$\big|\psi^{(N-\sum_{j=1}^{|\mathcal{K}_0|}b_j)}_{b_1\cdots b_{|\mathcal{K}_0|}}\big\ran$
are annihilated by $c_{\mb{k}_j}$ for all $j$ (i.e., these states lie in the space 
$\tilde{\mathcal{F}}$), and where
the coefficients $a_{b_1\cdots b_{|\mathcal{K}_0|}}$ satisfy
\beq
	\sum_{b_1,\dots,b_{|\mathcal{K}_0|}=0,1}|a_{b_1\cdots b_{|\mathcal{K}_0|}}|^2=1\ .
\eeq
As in the
one-dimensional case, some of the coefficients $a_{b_1\cdots b_{|\mathcal{K}_0|}}$ may be zero depending 
on the specific value of $N$, and all of these coefficients can be non-zero for 
$N\leq |\Lambda|-|\mathcal{K}_0|$.

As in the $D=1$ case, we can use this expression for $|\psi^{(N)}_0\ran$ 
and the fact that $[H,n_{\mb{k}_j}]=0$ for all $j$ to obtain the lower bound
\beq
	E^{(N)}_0 \geq \sum_{j=1}^{|\mathcal{K}_0|}h_j \ep_{\mb{k}_j}+\tilde{E}^{(M^*)}_0\ .
\eeq
We then define the trial state
\beq
	|\psi_t\ran= (c^{\dg}_{\mb{k}_1})^{h_1}\cdots (c^{\dg}_{\mb{k}_{|\mathcal{K}_0|}})^{h_{|\mathcal{K}_0|}}|\td{\psi}^{(M^*)}_0\ran\ ,
\eeq
where $|\td{\psi}^{(M^*)}_0\ran$ is the ground state of $\tilde{H}$ in the $M^*$-particle 
sector of $\tilde{\mathcal{F}}$ (or one of the ground states if $\tilde{H}$ has a ground 
state degeneracy in that sector). This
trial state has a particle number equal to $M^{**}$, where now
\beq
	M^{**}= M^* + \sum_{j=1}^{|\mathcal{K}_0|}h_j\ ,
\eeq
and the energy of this state is given by
\beq
	E_t= \lan\psi_t|H|\psi_t\ran=\sum_{j=1}^{|\mathcal{K}_0|}h_j \ep_{\mb{k}_j}+\tilde{E}^{(M^*)}_0\ .
\eeq
Using the same variational arguments from the previous subsection, 
we again find that $E^{(M^{**})}_0=E_t$ and that
$E^{(N)}_0 \geq E^{(M^{**})}_0$ for all $N$.
Finally, we can again apply our assumption of a non-zero parity gap 
$\Delta_0$, and the result of Lemma 1, to find that
\beq
	\mathcal{P}_0= (-1)^{M^{**}}= (-1)^{\sum_{j=1}^{|\mathcal{K}_0|}h_j}= \prod_{j=1}^{|\mathcal{K}_0|}s_j = \prod_{\mb{k}\in\mathcal{K}_0}s_{\mb{k}}\ .
\eeq 
This completes the proof of Theorem 1 for a general spatial dimension $D\geq 1$.

\section{Discussion and Conclusion}
\label{sec:con} 

We have proven that the mean-field approximation correctly predicts the value of 
the $\mathbb{Z}_2$ topological invariant $\nu=\mathcal{P}_0\mathcal{P}_{\pi}$ for \emph{any} 
translation invariant Hamiltonian that is adiabatically connected to a gapped pairing model of the form 
\eqref{eq:Ham}. We emphasize that this is a large family of models that is likely to contain many 
realistic models with a superconducting ground state. 
Our rigorous results give strong evidence that the mean-field approach is 
reliable, at least for the calculation of bulk topological invariants. As a topic for future work,
we propose to search for evidence of Majorana-like excitations in pairing
models with interfaces or boundaries, as our results strongly suggest that some kind of interesting gapless
excitations should appear at the boundary between two pairing models with opposite values of $\nu$.
Such a study would also have a direct impact on future experiments on TSCs, as the 
number-conserving pairing models are (presumably) a better description of the true experimental
situation than the mean-field models.

\acknowledgments 

We thank M. Levin for many discussions on this topic and for giving many
helpful comments on the first draft of this paper. We also acknowledge the support of the Kadanoff Center 
for Theoretical Physics at the University of Chicago. This work was supported by the Simons 
Collaboration on Ultra-Quantum Matter, which is a grant from the Simons Foundation (651440).

\appendix

\section{Reflection positivity of the pairing Hamiltonian $\tilde{H}$}
\label{app:RP}

In this appendix we introduce \emph{reflection positivity} in the setting relevant
for our work, and then we show that
the pairing Hamiltonian $\tilde{H}$ (Eq.~\eqref{eq:td-Ham} of the main text) is indeed reflection positive.
An interested reader should then be able to consult 
Ref.~\onlinecite{TT} to see how this property is used to prove Eq.~\eqref{eq:RP-result} of the main text.
Although the term ``reflection positivity'' has slightly different meanings in different areas of physics
(e.g., in quantum field theory, statistical mechanics, quantum mechanics, etc.), in our 
setting reflection positivity is simply the statement that the Hamiltonian of the system takes a certain 
special form. Some useful references for reflection positivity in this sense are 
Refs.~\onlinecite{KLS,lieb-hubbard} and Appendix 2 of Ref.~\onlinecite{LN}.
The special form of a reflection positive Hamiltonian is extremely useful because it 
allows for the derivation of various inequalities that can be used to prove many things about the model.

We first describe the general form of a reflection positive Hamiltonian. We start
with a Hilbert space $\mathcal{H}$. A reflection positive Hamiltonian acts on the tensor product space
$\mathcal{H}\otimes\mathcal{H}$ and takes a specific form that we now describe.
Let $A$ be a Hermitian operator on $\mathcal{H}$, let $\mathcal{J}$ be some
index set, and let $B_J$, for all $J\in\mathcal{J}$, be a set of \emph{real} operators on $\mathcal{H}$
(i.e., the matrix elements of the $B_J$ are real in a certain preferred basis for $\mathcal{H}$). 
In addition, let $g_J\geq 0$ be a set of nonnegative real coefficients labeled by $J\in\mathcal{J}$, and
assume that $\sum_{J\in\mathcal{J}}g_J B_J\otimes B_J$ is symmetric as an 
operator on $\mathcal{H}\otimes\mathcal{H}$. In terms of these ingredients, a reflection
positive Hamiltonian $H_{\text{RP}}$ acting on $\mathcal{H}\otimes\mathcal{H}$ takes the form
\beq
	H_{\text{RP}}= A\otimes\mathbb{I}+\mathbb{I}\otimes A - \sum_{J} g_J B_J\otimes B_J\ . \label{eq:RP}
\eeq
The minus sign on the term with the $B_J$ (and the nonnegativity of the $g_J$)
is crucial for reflection positivity. When the two factors of $\mathcal{H}$
represent spin-up and spin-down degrees of freedom, a system with a Hamiltonian of the
form $H_{\text{RP}}$ is said to possess Lieb's \emph{spin reflection positivity}.~\cite{lieb-hubbard}

We now show that the pairing Hamiltonian $\tilde{H}$ from 
Eq.~\eqref{eq:td-Ham} of the main text is indeed a Hamiltonian of the form \eqref{eq:RP}. 
We consider $\tilde{H}$ acting within the Fock space
$\mathcal{F}_{\mathcal{K}_{+}\cup\mathcal{K}_{-}}$ for the fermions labeled
by wave vectors $\mb{k}$ in the set $\mathcal{K}_{+}\cup\mathcal{K}_{-}$ (note that this Fock space is
equivalent to the space $\tilde{\mathcal{F}}$ that we defined in Eq.~\eqref{eq:Fock} of the main text).
When written in terms of the spinful fermion operators $c_{\mb{k},\sigma}$
that we introduced in the main text (with $\mb{k}\in\mathcal{K}_{+}$ and $\sigma\in\{\up,\down\}$),
$\tilde{H}$ takes the form
\beq
	\tilde{H}= \sum_{\mb{k}\in\mathcal{K}_{+}}\sum_{\sigma=\up,\down}\ep_{\mb{k}} n_{\mb{k},\sigma} - \sum_{\mb{k},\mb{k}'\in \mathcal{K}_{+}} |\eta_{\mb{k}}||\eta_{\mb{k}'}| c^{\dg}_{\mb{k},\up} c_{\mb{k}',\up} c^{\dg}_{\mb{k},\down}c_{\mb{k}',\down}\ .
\eeq
To proceed, we define separate number operators for the spin-up and spin-down fermions, 
$\mathcal{N}_{\sigma}= \sum_{\mb{k}\in\mathcal{K}_{+}}n_{\mb{k},\sigma}$. 
We then use these number operators to 
define new spinful fermion operators $C_{\mb{k},\sigma}$ as follows. For spin-up we set
$C_{\mb{k},\up}$ equal to $c_{\mb{k},\up}$,
\beq
	C_{\mb{k},\up}= c_{\mb{k},\up}\ .
\eeq
On the other hand, for spin-down we define $C_{\mb{k},\down}$ via 
\beq
	C_{\mb{k},\down}= (-1)^{\mathcal{N}_{\up}}c_{\mb{k},\down}\ .
\eeq
With this definition we still find that $\{C_{\mb{k},\sigma},C^{\dg}_{\mb{k}',\sigma}\}=\delta_{\mb{k}\mb{k}'}$ and $\{C_{\mb{k},\sigma},C_{\mb{k}',\sigma}\}=0$ for operators with the same spin, 
but now we find that operators with opposite spins \emph{commute} instead of anticommute, for example
\beq
	[C_{\mb{k},\up},C_{\mb{k}',\down}]= [C_{\mb{k},\up},C^{\dg}_{\mb{k}',\down}]=0\ .
\eeq
On the other hand, $\tilde{H}$ takes the exact same form when written in terms of the
new fermion operators, 
\beq
	\tilde{H}= \sum_{\mb{k}\in\mathcal{K}_{+}}\sum_{\sigma=\up,\down}\ep_{\mb{k}} N_{\mb{k},\sigma} - \sum_{\mb{k},\mb{k}'\in \mathcal{K}_{+}} |\eta_{\mb{k}}||\eta_{\mb{k}'}| C^{\dg}_{\mb{k},\up} C_{\mb{k}',\up} C^{\dg}_{\mb{k},\down}C_{\mb{k}',\down}\ ,
\eeq
where we defined $N_{\mb{k},\sigma}= C^{\dg}_{\mb{k},\sigma}C_{\mb{k},\sigma}$.

We are now ready to show that $\tilde{H}$ is reflection positive. 
Since the spin-up fermions $C_{\mb{k},\up}$ commute with the spin-down fermions $C_{\mb{k},\down}$, we can 
now regard the Hamiltonian $\tilde{H}$ as acting on the tensor product space
\beq
	\mathcal{F}_{\mathcal{K}_{+}}\otimes\mathcal{F}_{\mathcal{K}_{+}} \ , \nnb 
\eeq
where $\mathcal{F}_{\mathcal{K}_{+}}$ is the Fock space for a single set of fermions 
$C_{\mb{k}}$ labeled only by wave vectors $\mb{k}$ in $\mathcal{K}_{+}$ (no additional spin index). 
We now see that $\tilde{H}$ can be written in the reflection positive form \eqref{eq:RP}, if we make the
following identifications. First, the Hilbert space $\mathcal{H}$ is equal to the Fock space
$\mathcal{F}_{\mathcal{K}_{+}}$ for the fermions operators $C_{\mb{k}}$. 
Next, the Hermitian operator $A$ is given by
\beq
	A = \sum_{\mb{k}\in\mathcal{K}_{+}}\ep_{\mb{k}} N_{\mb{k}}
\eeq
where $N_{\mb{k}}= C^{\dg}_{\mb{k}}C_{\mb{k}}$. Finally, the abstract index $J$ is identified
with pairs $(\mb{k},\mb{k}')$ of wavevectors in $\mathcal{K}_{+}$, $J= (\mb{k},\mb{k}')$, and
the operators $B_J$ and coefficients $g_J$ are given by
\beq
	B_J= B_{(\mb{k},\mb{k}')} = C^{\dg}_{\mb{k}} C_{\mb{k}'}
\eeq
and
\beq
	g_J = g_{(\mb{k},\mb{k}')}= |\eta_{\mb{k}}||\eta_{\mb{k}'}|\ .
\eeq 
One can also check that the $B_J$ operators have real matrix elements in the occupation number
basis for $\mathcal{F}_{\mathcal{K}_{+}}$ in which all the number operators $N_{\mb{k}}$ are diagonal.

This completes our introduction to reflection positivity and our demonstration that the pairing Hamiltonian
$\tilde{H}$ has this property. This information should allow any interested reader to consult
Ref.~\onlinecite{TT} to see how reflection positivity is used to prove Eq.~\eqref{eq:RP-result} of the main text.


%

\end{document}